\documentclass[runningheads]{llncs}
\usepackage{graphicx}
\usepackage{hyperref}
\usepackage[hyphenbreaks]{breakurl}
\usepackage{booktabs}
\usepackage{stmaryrd}
\usepackage[T1]{fontenc}
\usepackage{cite}

\usepackage[ruled,vlined,linesnumbered]{algorithm2e}
\usepackage{algpseudocode}
\usepackage{graphicx}
\usepackage[export]{adjustbox}
\usepackage{subcaption}
\usepackage{float}

\usepackage[table]{xcolor}
\usepackage{amssymb,amsmath}

\begin{document}
\title{Performance Evaluation of GPS Trajectory Rasterization Methods}
%
%
\author{Necip E. Gengeç\inst{1}\orcidID{0000-0001-7857-8986} \and
Ergin Tarı\inst{2}\orcidID{0000-0002-9873-6854}}
\authorrunning{N.E. Gengeç, E. Tarı}
%
\institute{Graduate School of Engineering and Technology, Istanbul Technical University, Istanbul, Turkey
\email{gengec@itu.edu.tr} \and
Department of Geomatics Engineering, Istanbul Technical University, Istanbul, Turkey\\
\email{tari@itu.edu.tr}}
\maketitle              
\begin{abstract}
The availability of the Global Positioning System (GPS) trajectory data is increasing along with the availability of different GPS receivers and with the increasing use of various mobility services. GPS trajectory is an important data source which is used in traffic density detection, transport mode detection, mapping data inferences with the use of different methods such as image processing and machine learning methods. While the data size increases, efficient representation of this type of data is becoming difficult to be used in these methods. A common approach is the representation of GPS trajectory information such as average speed, bearing, etc. in raster image form and applying analysis methods. In this study, we evaluate GPS trajectory data rasterization using the spatial join functions of QGIS, PostGIS+QGIS, and our iterative spatial structured grid aggregation implementation coded in the Python programming language. Our implementation is also parallelizable, and this parallelization is also included as the fourth method. According to the results of experiment carried out with an example GPS trajectory dataset, QGIS method and PostGIS+QGIS method showed relatively low performance with respect to our method using the metric of total processing time. PostGIS+QGIS method achieved the best results for spatial join though its total performance decreased quickly while test area size increases. On the other hand, both of our methods' performances decrease directly proportional to GPS point. And our methods' performance can be increased proportional to the increase with the number of processor cores and/or with multiple computing clusters.

\end{abstract}

\section{Introduction}
\label{sec:introduction}
Availability of digital data is increasing with the increase of sensor device connectivity and with the decrease in data storage area costs. The availability of spatial data, the data that are having spatial components, is also increasing. Spatial data is collected and stored mostly with the use of Global Positioning System (GPS) receivers or other devices that are equipped with GPS units such as smart phones, navigation devices etc.

Collected GPS data with receivers is also varying. One type of data collected with GPS devices is called GPS trajectories and it is the collection of consecutive GPS locations during the travel time of a moving body \cite{Zheng2009_geolife2}. In addition to GPS locations, additional information such as timestamp, speed, bearing of the movement, acceleration/deceleration can be recorded with GPS trajectory and/or can be derived from one another.

\begin{figure}[ht]
    \centering
    \includegraphics[width=\linewidth, frame]{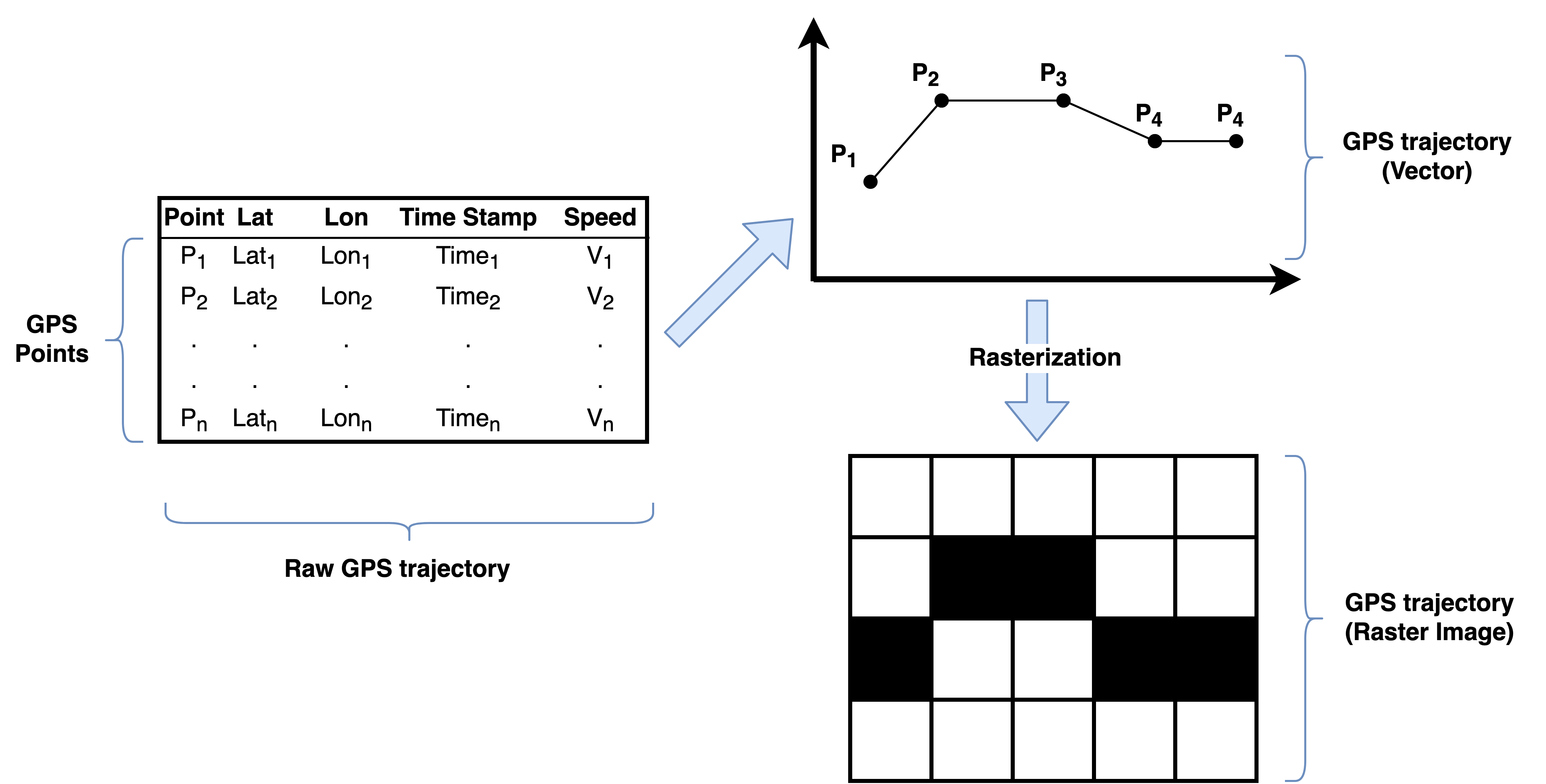}
    \caption{A simple GPS trajectory and its rasterization.}
    \label{fig:rasterization_definition}
\end{figure}

GPS trajectories are used in studies focusing on mapping data inference \cite{he2018roadrunner, Karagiorgou_2012, 10.1007/978-3-642-33090-2_7}, traffic density detection \cite{traffic4cast} and transportation mode detection \cite{DABIRI2018360, trajectorynet_reference}. In these examples from literature GPS trajectories are used as is or represented in a generalized form such as embedding attributes into predetermined feature classes or converting GPS trajectories into raster images (Figure \ref{fig:rasterization_definition}) that are representing certain attributes (GPS point frequency, transportation mode) or their attributes' aggregation (average speed, maximum speed, average bearing). After the representation of GPS trajectories with embedding or rasterization, different data analysis methods can be applied to these derived data.

{}

In addition, research on GPS trajectories as GPS trajectories being the only data source, GPS trajectories fusion with satellite or aerial imagery is a new area \cite{sun2019leveraging}. In the context of data fusion of GPS trajectories with satellite imagery, GPS trajectories are rasterized, and it is important to obtain one to one pixel match between rasterized GPS trajectories and satellite imagery to carry out the analysis accurately.

On the other hand, due to the size of GPS trajectory data, the rasterization process can be time consuming while the data size and work area increases. This issue is not limited only to GPS trajectory rasterization or aggregation, similar research domains such as spatial social media data analysis and other domains that are dealing with high volume point data.

In this contribution, we address the rasterization of GPS trajectories using open source Geographic Information System (GIS) tools and an algorithm coded using the Python programming language. These tools are evaluated according to their performances with an experiment which has the goal to rasterize multiple attributes of given GPS trajectories. Evaluation carried out only on the performance results of approaches for three tools in the same architecture is presented, no philosophical discussion is carried out yet.

To the best of our knowledge, this study is the only study comparing multiple open source tools and algorithms to understand their performance for aggregation of the big point data to structured grids and their rasterization. There has been multiple research for general performance comparison of QGIS with respect to GIS software like ArcGIS \cite{friedrich2014comparison}. The parallelization is one of the options implemented in this study. There is various research on parallelization for GIS applications spreading from implementing big data tools into GIS software \cite{arcgis_hadoop, arcgis_hadoop2, grass_hpc_ref} to adaption of cluster based, distributed big data tools into GIS domain \cite{spatial_hadoop_ref, hadoop_gis_ref}. Also, there are significant research which discuss CPU and GPU acceleration, their special applications in GIS and achieved performance improvement \cite{gpu_cpu__zhang2012_1, gpu_cpu_zhang2012_2, gpu_cpu_zacharatou2017}. Although previous researches may contribute to various future research directions combined with our research results, these researches neither focus on the point to structured grid aggregation and rasterization nor provide a performance comparison with respect to the widely used open source GIS tools.

\section{Tools and Rasterization Process Flow}
\label{sec:tools_processflow}
As in raw form, GPS trajectory data is a vector data while the aim of this research is to represent this data in raster image format. Because of these dependencies, tools that are required should be able to handle both vector and raster image data. Within the multiple open source GIS tools that are freely available, QGIS \cite{qgis_ref} and PostGIS \cite{postgis_ref} are used for the rasterization of GPS trajectories in this study since they are widely adopted in GIS field thanks to their robustness and abilities. 

In Section \ref{sec:rasterization_qgis} and \ref{sec:rasterization_postgis}, the rasterization abilities of QGIS and PostGIS will be examined with respect to given input data (work area boundaries, output pixel size, GPS trajectory data) and expected output rasterized GPS trajectory layers (frequency, average speed and maximum speed). In Section \ref{sec:rasterization_python}, our implementation will be explained. Finally, in Section \ref{sec:process_flow_sub}, the process flow of methods will be summarized.

\subsection{Rasterization with QGIS}
\label{sec:rasterization_qgis}
As a GIS software, QGIS has different functions, tools and plugins for data analysis and data conversion. \textit{Rasterize (Vector to Raster)} is one of these tools offered within QGIS. This QGIS tool acts as a user interface, collects the user inputs and runs \textit{gdal\_rasterize} tool at the background. This tool is able to get an input data and burn the pixel values that are stored in the preferred attribute field of input vector data within the predefined outer boundaries. Although QGIS has the rasterize tool, this tool is only able to rasterize given values but cannot aggregate multiple values of the same attribute in the given pixels. Though, it is possible to represent pixels in vector form (structured grid) and achieve the required aggregation with spatial join tool of QGIS. After the aggregation of GPS trajectories into the structured grid, it is possible to rasterize the aggregated attributes into raster image data.

\subsection{Rasterization with PostGIS}
\label{sec:rasterization_postgis}
PostGIS is the spatial database add-on for the PostgreSQL \cite{postgresql_ref} database management system. PostGIS is able to store and analyze spatial data in vector and raster image form that is stored in a PostgreSQL database. PostGIS has \textit{ST\_AsRaster} tool for similar to QGIS which is accepting the input though producing only given attribute values. On the other hand, similar to QGIS, it is possible to aggregate one vector layer into another using Structured Query Language (SQL) statements. Even though PostGIS does provide rasterize functionality, it is also possible to connect QGIS to PostgreSQL database and rasterize the output data that is created with PostGIS via QGIS.

\subsection{Rasterization with Python}
\label{sec:rasterization_python}
Python is a general-purpose programming language which has many internal and external libraries such as data science libraries (Pandas) and geospatial computation (GDAL, pyproj) libraries. With the use of these libraries, it is possible to analyze GPS trajectories. 

To achieve the required raster images, our own Python method was created (Algorithm \ref{alg:python_script}). This method gets the GPS points, coordinates of work area boundary and pixel size as inputs and calculates the raster matrix. Unlike PostGIS and QGIS, the Python method makes use of the structured grid definition (work area boundary and pixel size) of a raster image and carries out the calculation of outputs without creating a vector grid. The method determines the row and column of the pixel where each GPS point is contained. Following, it aggregates the required feature values and assigns the output raster image matrix values according to previously determined rows and columns.

The most computation intensive part of the Algorithm \ref{alg:python_script} is the $for$ loop shown between row numbers 5 to 9. Python gives the ability to parallelize with the use of additional libraries such as Dask \cite{dask_ref} and Swifter \cite{swifter_ref}. Swifter library uses Dask library at its backend. It is able to provide the processing time information and provides user the ability to choose parallel or normal computation options easily. Due to these features Swifter library is used for the experiments of this study.

\begin{center}
\begin{algorithm}[H]
\SetAlgoLined
\DontPrintSemicolon
\KwData{$P=\{p_1, p_2, p_3,....,p_n\}$ where each $p_i$ contains latitude ($p_{i,lat}$), longitude($p_{i,lon}$), speed ($p_{i,speed}$).\newline Output raster image top-left corner coordinate ($X,Y$) and pixel size ($px$).\;}
\KwResult{Output images $Image_{count}$, $Image_{speed-avg}$, $Image_{speed-max}$ in matrix form. }
    \Begin{
        \tcc{Convert input coordinates into projected cartesian coordinate system.}
        
        \SetKwProg{Def}{def}{:}{}
        \Def{transformCoordinates($P_{lat}$, $P_{lon}$)}{
        Transform WGS84 to projected WGS84
        }{return $P_X$, $P_Y$\;}
        
        \tcc{Determining row and column of each GPS point within the output raster.}
        \ForEach{$p_i \in P$}{
            $p_{row} = (p_{i,X} - (p_{i,X}\bmod px) - X) / px$\;
            $p_{column} = (p_{i,Y} - (p_{i,Y}\bmod px) - Y) / px$\;
            $p_i \longleftarrow p_{row}, p_{column}$\;
        }
        \SetKwProg{Def}{def}{:}{}
        \tcc{Aggregate GPS count, average and maximum speed values with grouping by row and column number.}
        \Def{aggregateValues($P_{row}$, $P_{column}$, $P_{speed}$)}{
            $P_{count}^{'} \longleftarrow$ count of records having same $p_{row}, p_{column}$ where $p \in P$\;
            $P_{speed-avg}^{'} \longleftarrow$ average of $p_{i,speed}$ having same $p_{row}, p_{column}$ where $p \in P$\;
            $P_{speed-max}^{'} \longleftarrow$ maximum of $p_{i,speed}$ having same $p_{row}, p_{column}$ where $p \in P$\;
        }{return $P ^ {'}$\;}
        \tcc{Assign pixel values by row and column of output images.}
        \ForEach{$p_i^{'} \in P^{'}$}{
            $Image_{count}[p_{i,row}^{'}][p_{i,column}^{'}] \longleftarrow p_{i,count}^{'}$\;
            $Image_{speed-avg}[p_{i,row}^{'}][p_{i,column}^{'}] \longleftarrow p_{i,speed-avg}^{'}$\;
            $Image_{speed-max}[p_{i,row}^{'}][p_{i,column}^{'}] \longleftarrow p_{i,speed-max}^{'}$\;
        }
    }
 \caption{Spatial join with Python.}
 \label{alg:python_script}
\end{algorithm}
\end{center}

\subsection{Summary of Methods}
\label{sec:process_flow_sub}
\begin{figure}[ht]
    \centering
    \includegraphics[width=\linewidth]{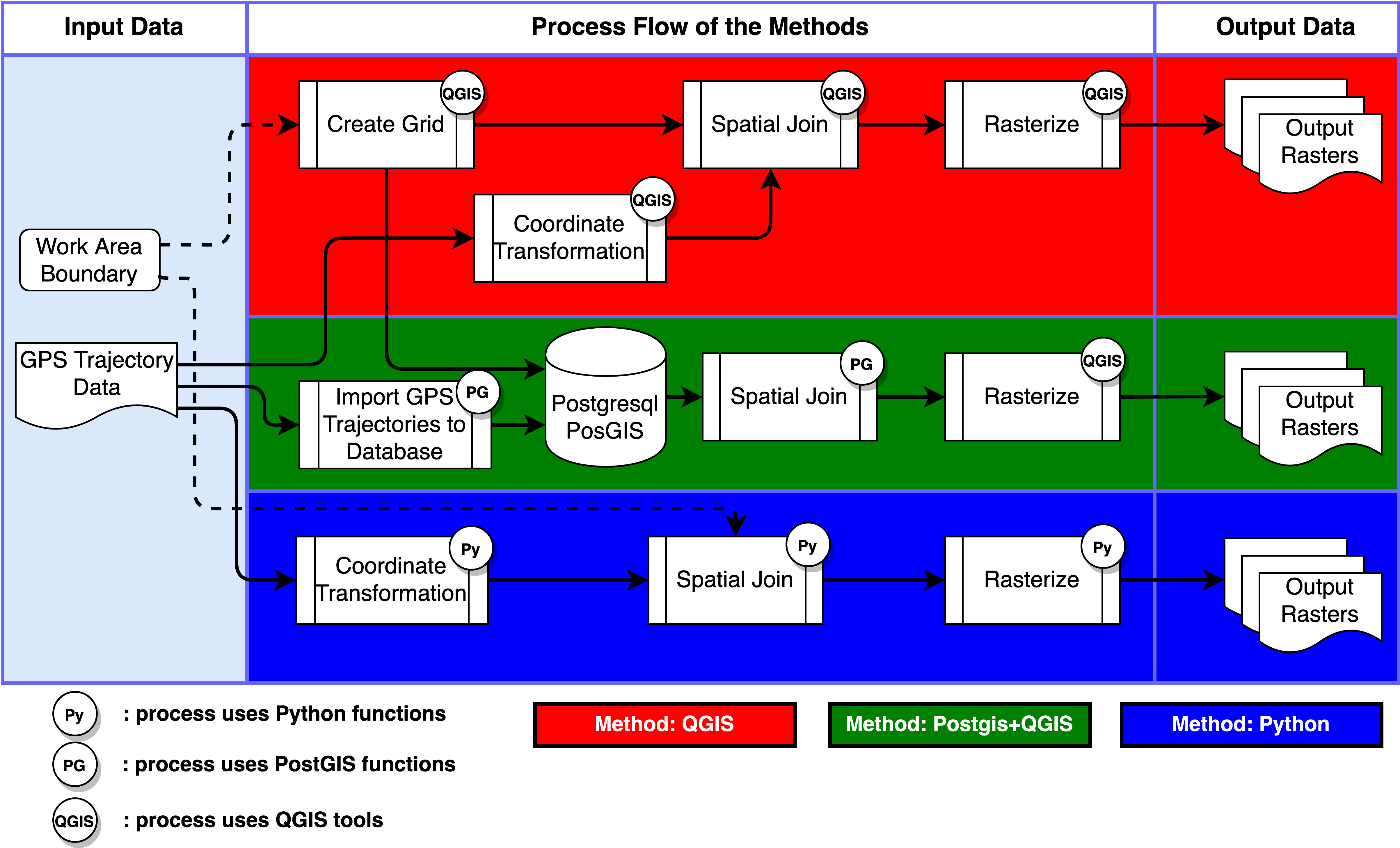}
    \caption{Process flow of the methods; QGIS, PostGIS+QGIS and Python (Python process flow is identical for both Python and Python (Parallel) methods).}
    \label{fig:process_flow}
\end{figure}

According to the spatial data processing and rasterization abilities of QGIS, PostGIS and Python the process flow of the four methods was determined as in the Figure \ref{fig:process_flow}.

The QGIS method creates a vector grid and transforms coordinates of GPS points into the coordinate system of the required output raster image. After creation of the vector grid and coordinate transformation, these are joined spatially. Finally, output of the spatial join is rasterized.

{}

The second method is called PostGIS+QGIS because this method benefits from both PostGIS and QGIS. This method creates the vector grid with the use of QGIS. Also, GPS trajectories are required to be imported into PostgreSQL database which PostGIS is operating on. After import and grid creation, PostGIS spatially joins both data. In this method, coordinate transformation is carried out along with the spatial join. Finally, QGIS is used to rasterize the output of the spatial join.

The algorithm for the Python method is explained in Algorithm \ref{alg:python_script}. This algorithm gets the GPS trajectory data directly using Python libraries and applies the coordinate transformation. After the transformation, our method calculates the output raster image matrix  without the need of a vector grid and saves the output raster to the disk.

\section{Setup of the Experiments}
\label{sec:experimental_setup}
The methods defined in Section \ref{sec:tools_processflow} are evaluated with the use of MTL-Trajet dataset \cite{mtl_traject_reference}. This dataset consists of GPS trajectories collected in 2016 around Montreal, Canada. Raw data contains GPS point locations in the WGS84 Datum and timestamps. Speed of the moving objects are calculated using time difference and geodesic distances of consecutive GPS points with Geopy package \cite{geopy_ref}. These values added as an attribute to the raw GPS trajectory data. Speed value accuracy is dependent to projection and calculation method, but speed accuracy is not the main focus of this study. Because of this, achieved speed values are well enough for performance evaluation of the methods.

\begin{figure}[H]
    \centering
    \includegraphics[width=0.65\linewidth, frame]{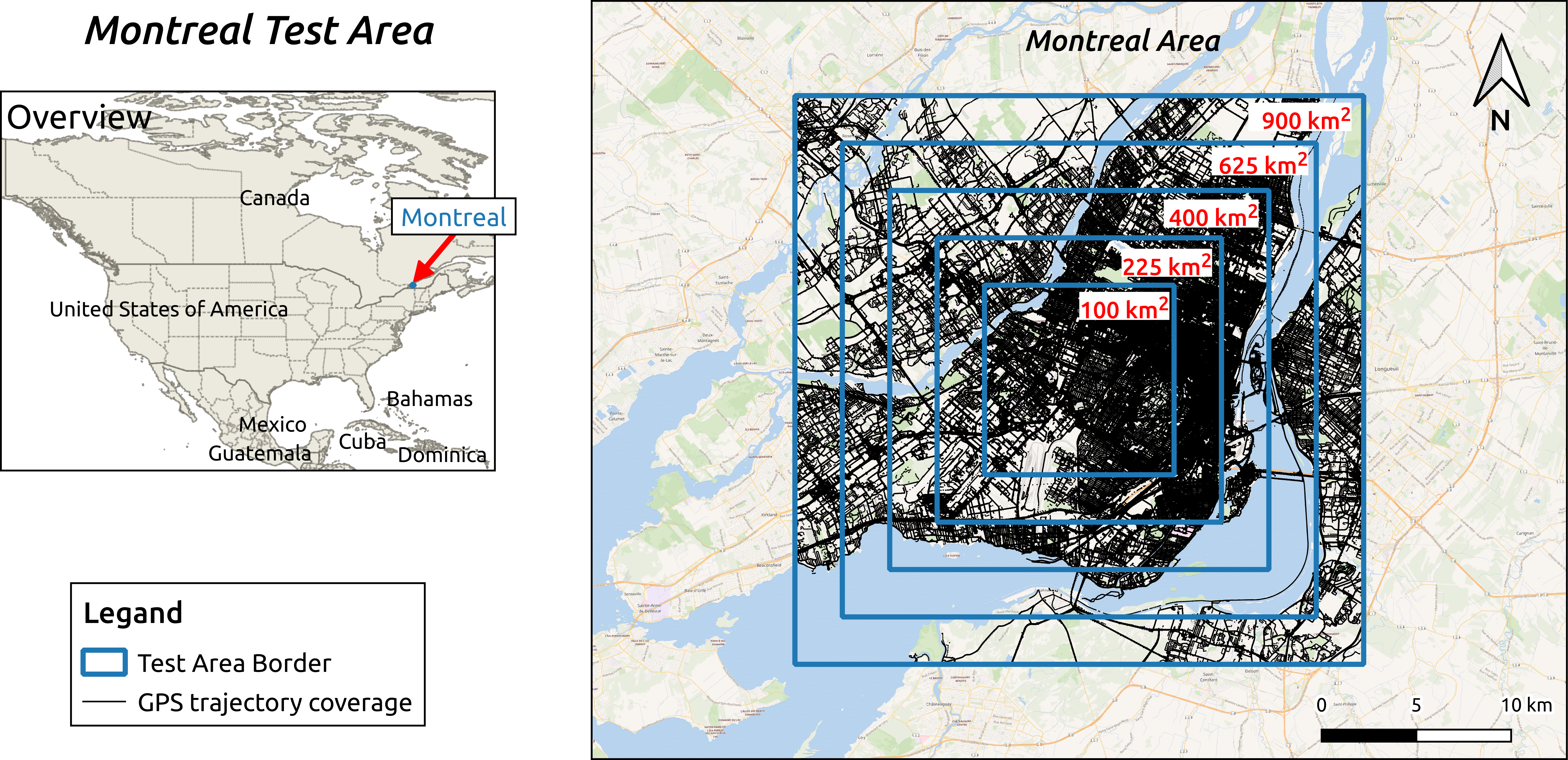}
    \caption{MTL-Trajet dataset and test areas boundaries.}
    \label{fig:test_areas}
\end{figure}

In the experiment, the target is to rasterize GPS point "frequency (count)", "average speed" and "maximum speed" raster images using QGIS, PostGIS+QGIS, Python and Python (Parallel) methods. In order to understand the dependencies of performance, the experiment is carried out with varying test area size and GPS point count. Figure \ref{fig:test_areas} shows the test area boundaries and Table \ref{tab:work_area_vs_gps_count} summarizes the corresponding GPS count that is used in the experiments. MTL-Trajet GPS point coverage is wider than the defined test areas. The GPS points that are outside of the test area are removed from main dataset to focus only on the performance of methods for given area. Though our Python and Python (Parallel) implementations are able to ignore the GPS points which are out of given area.

{}

\begin{table}[H]
\caption{GPS trajectory by test area that was used in the experiment.}
\label{tab:work_area_vs_gps_count}
\centering
  \begin{tabular}{| c | l |}
    \hline
    \textbf{Test Area Size ($km^2$)} & \textbf{Number of GPS points (million)}\\
    \hline
    100 & 5 \\
    \hline
    225 & 5, 10, 11\\
    \hline
    400 & 5, 10, 15 \\
    \hline
    625 & 5, 10, 15, 18 \\
    \hline
    900 & 5, 10, 15, 20 \\
    \hline
  \end{tabular}
\end{table}

As explained in Section \ref{sec:introduction}, the aim is to obtain a one to one pixel match with a given raster image. Usually the raster images that are widely used are provided in projected coordinate systems. In order to add this constraint to the experimental setup, unlike GPS trajectory data, test areas are created in projected WGS84 Datum using cartesian coordinates so that the pixel dimensions are defined in meters. According to this, GPS point coordinates must be transformed into projected WGS84 from geographic WGS84. 

The aim of our research for the rasterization is to represent GPS points in raster format to use in further analysis with additional satellite imagery. Because of this, the expected output should be aligned to the satellite imagery pixel resolution. Similar studies in literature use high resolution satellite images that are having pixel resolution around 1-5 meters \cite{Papadomanolaki2016, mnih2010learning, sun2019leveraging}. In order to align with the literature examples the output raster image pixel size set as 5 meters.

Total processing time ($t_{Total time}$) and spatial join time ($t_{Spatial Join}$) are preferred as the evaluation metric. Since the process flow of each method is different, their total processing time is also varying. 

Total processing time with QGIS method is calculated with 

\begin{equation}
t_{Total QGIS} = t_{Grid creation}^{QGIS} + t_{Coordinate Transform}^{QGIS} + t_{Spatial Join}^{QGIS} + t_{Rasterize}^{QGIS}
\end{equation}

where $t_{Grid creation}^{QGIS}$ grid creation time, $t_{Coordinate Transform}^{QGIS}$ coordinate transformation time, $t_{Spatial Join}^{QGIS}$ spatial join time, $t_{Rasterize}^{QGIS}$ rasterization time with QGIS.

Total processing time with PostGIS+QGIS method is calculated with

\begin{equation}
t_{Total PostGIS+QGIS} = t_{Grid creation}^{QGIS}+ t_{Spatial Join}^{PG} + t_{Rasterize}^{QGIS}
\end{equation}

where $t_{Grid creation}^{QGIS}$ grid creation time with QGIS, $t_{Spatial Join}^{PG}$ spatial join with PostGIS, $t_{Rasterize}^{QGIS}$ rasterization time with QGIS.

Total processing time with Python method is calculated with

\begin{equation}
t_{Total Python} = t_{Coordinate Transform}^{Py} + t_{Spatial Join}^{Py} + t_{Rasterize}^{Py}
\end{equation}

where $t_{Coordinate Transform}^{Py}$ coordinate transformation, $t_{Spatial Join}^{Py}$ spatial join and $t_{Rasterize}^{Py}$ rasterization using Python.

Experiments are carried out using a standard laptop which has an Intel Core I7-4600M CPU with four 2.90GHz cores, 16GB RAM and Ubuntu/Linux operating system. The QGIS and PostgreSQL/PostGIS are used with their default installation settings. Spatial indexes are used for GPS point and vector grid layers in PostGIS+QGIS method. In order to avoid delays caused by memory and processor usage of another software, minimum required software was kept open while running a method.

\section{Comparisons}
\label{sec:comparisons}
Experiments are carried out with the methods defined in Section \ref{sec:process_flow_sub} and with the setup defined in Section \ref{sec:experimental_setup}. Following the experiments, output raster images are compared with the use of raster calculation. All output raster images subtracted from remaining output images one by one for given test area. This comparison aims to determine if the created raster images of each method is identical or not. If compared raster images are identical, empty raster image expected as the result of raster image subtraction. According to the comparison of the outputs of each test area, the subtraction results were empty images which proves that each method achieved the same raster images as output.

\begin{figure}[H]
\captionsetup[subfigure]{justification=centering}
\begin{subfigure}{.5\textwidth}
  \centering
  \includegraphics[width=\linewidth]{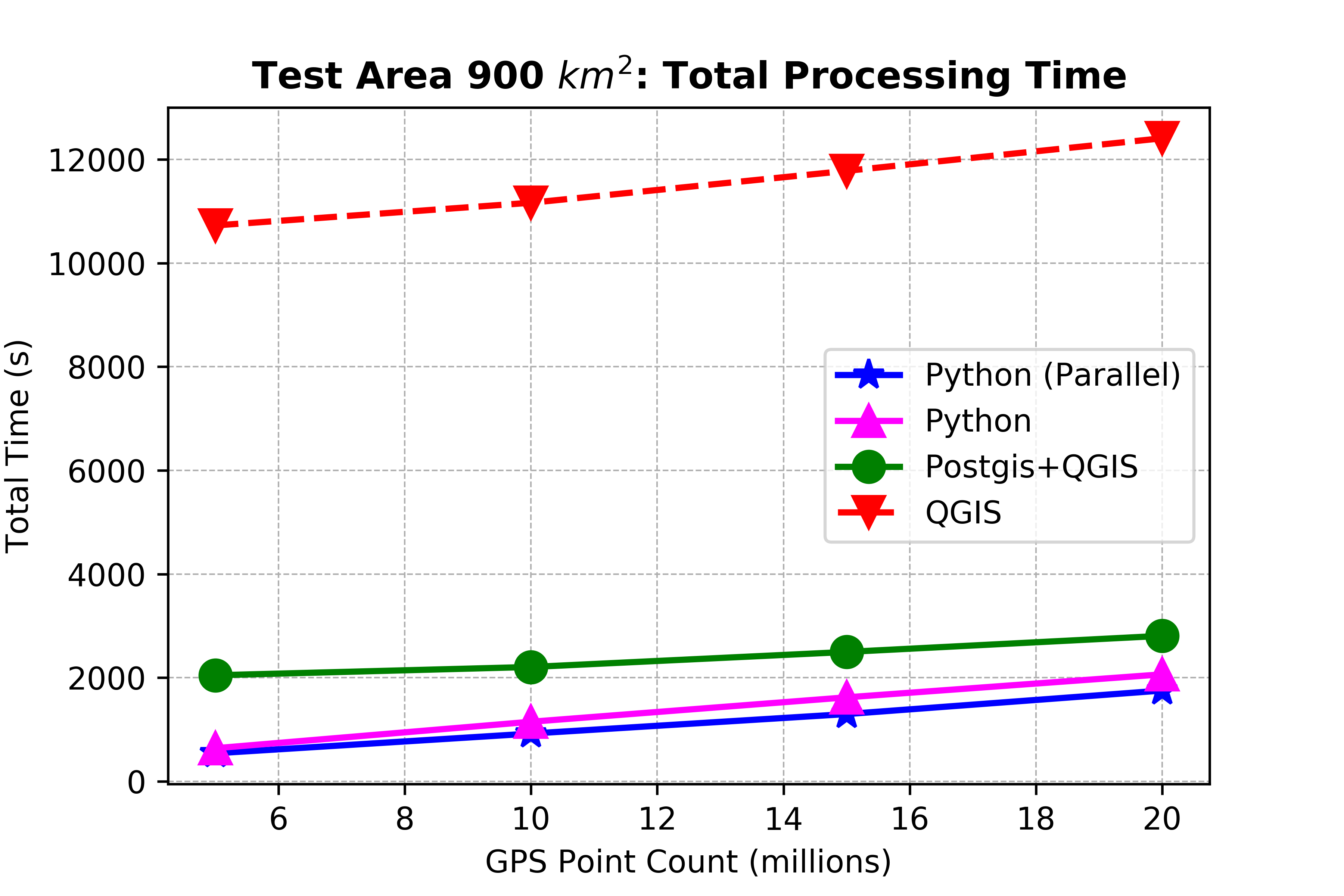}
  \caption{}
  \label{fig:900_all_qgis}
\end{subfigure}%
\begin{subfigure}{.5\textwidth}
  \includegraphics[width=\linewidth]{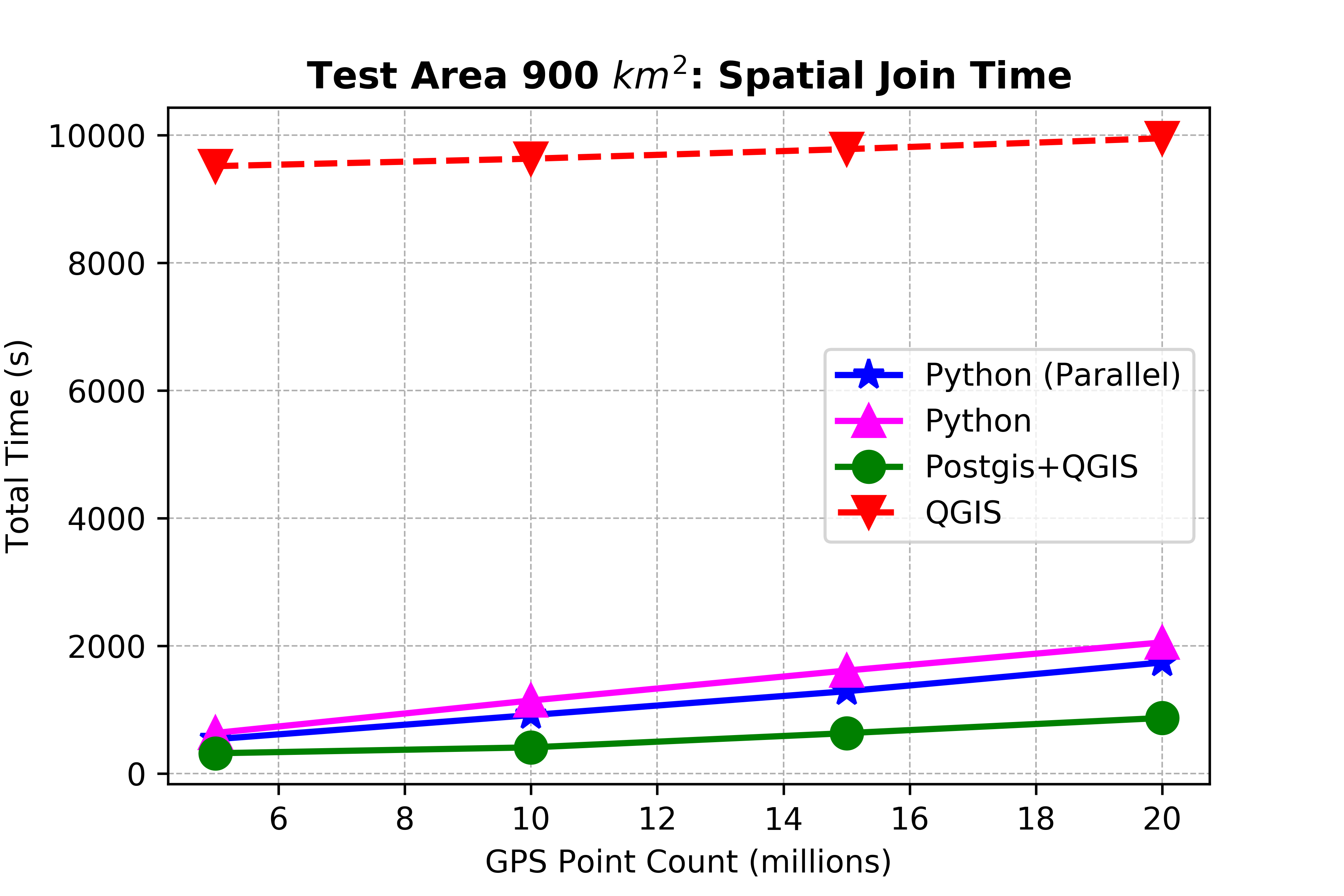}
  \centering
  \caption{}
  \label{fig:900_all_sptl_qgis}
\end{subfigure}
\caption{\textbf{(a)} comparison of total processing time and \textbf{(b)} comparison of spatial join time for 900 $km^2$ test area.}\label{fig:gps_comparison_with_qgis}
\end{figure}

\begin{figure}[H]
\captionsetup[subfigure]{justification=centering}
\begin{subfigure}{.5\textwidth}
  \centering
  \includegraphics[width=.90\linewidth]{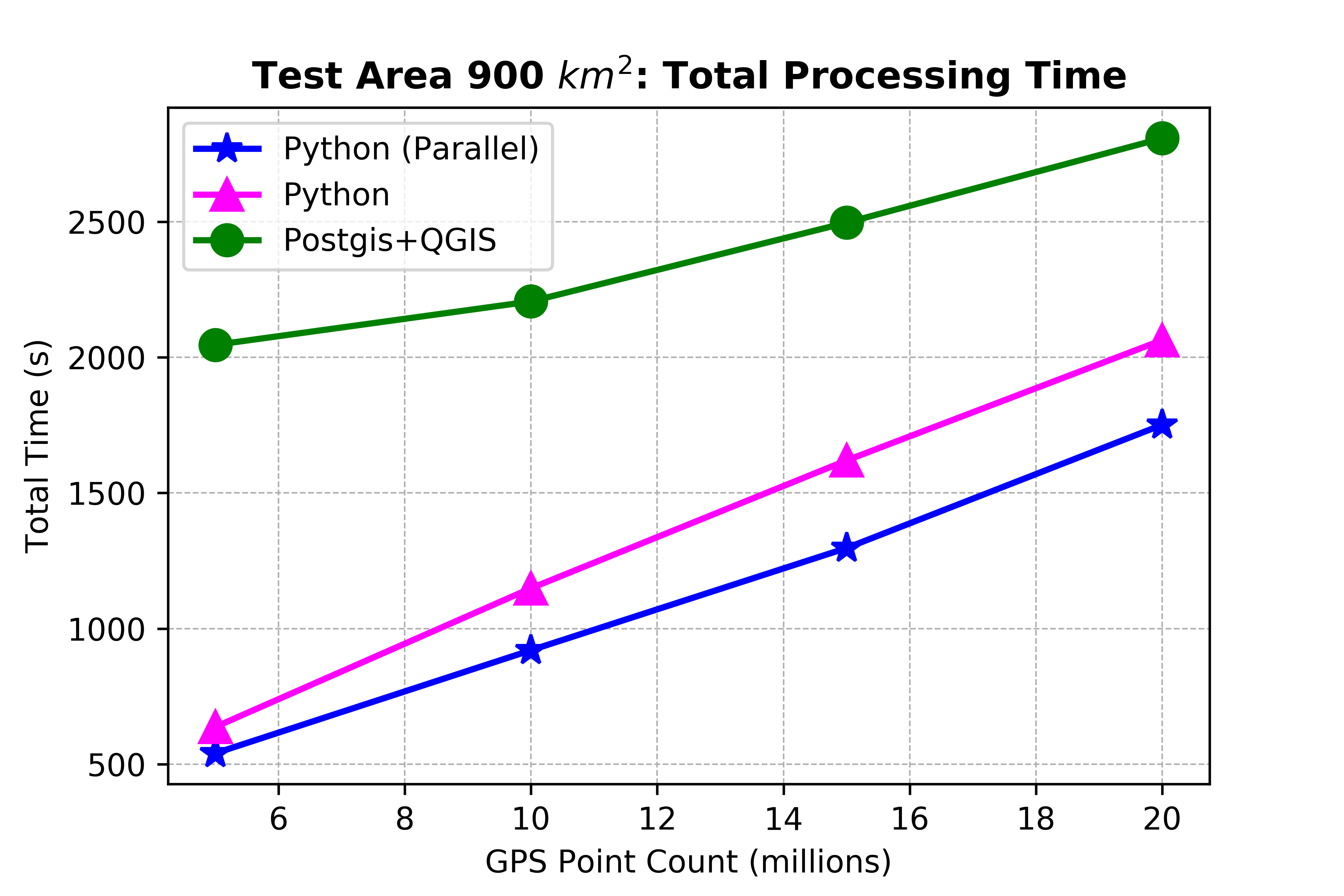}
  \caption{}
  \label{fig:900_all}
\end{subfigure}%
\begin{subfigure}{.5\textwidth}
  \includegraphics[width=.90\linewidth]{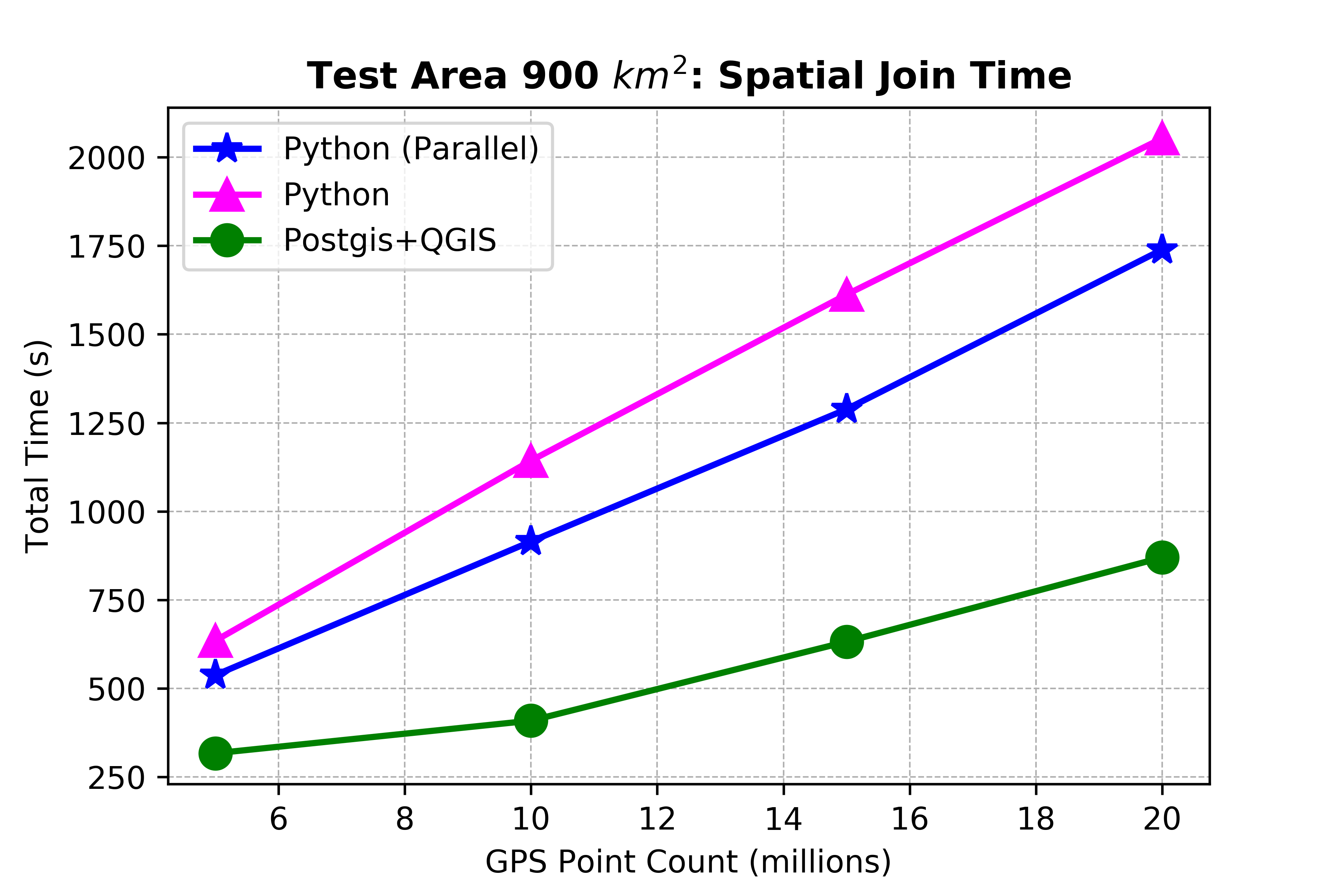}
  \centering
  \caption{}
  \label{fig:s900_all_sptl}
\end{subfigure}
\begin{subfigure}{.5\textwidth}
  \centering
  \includegraphics[width=.90\linewidth]{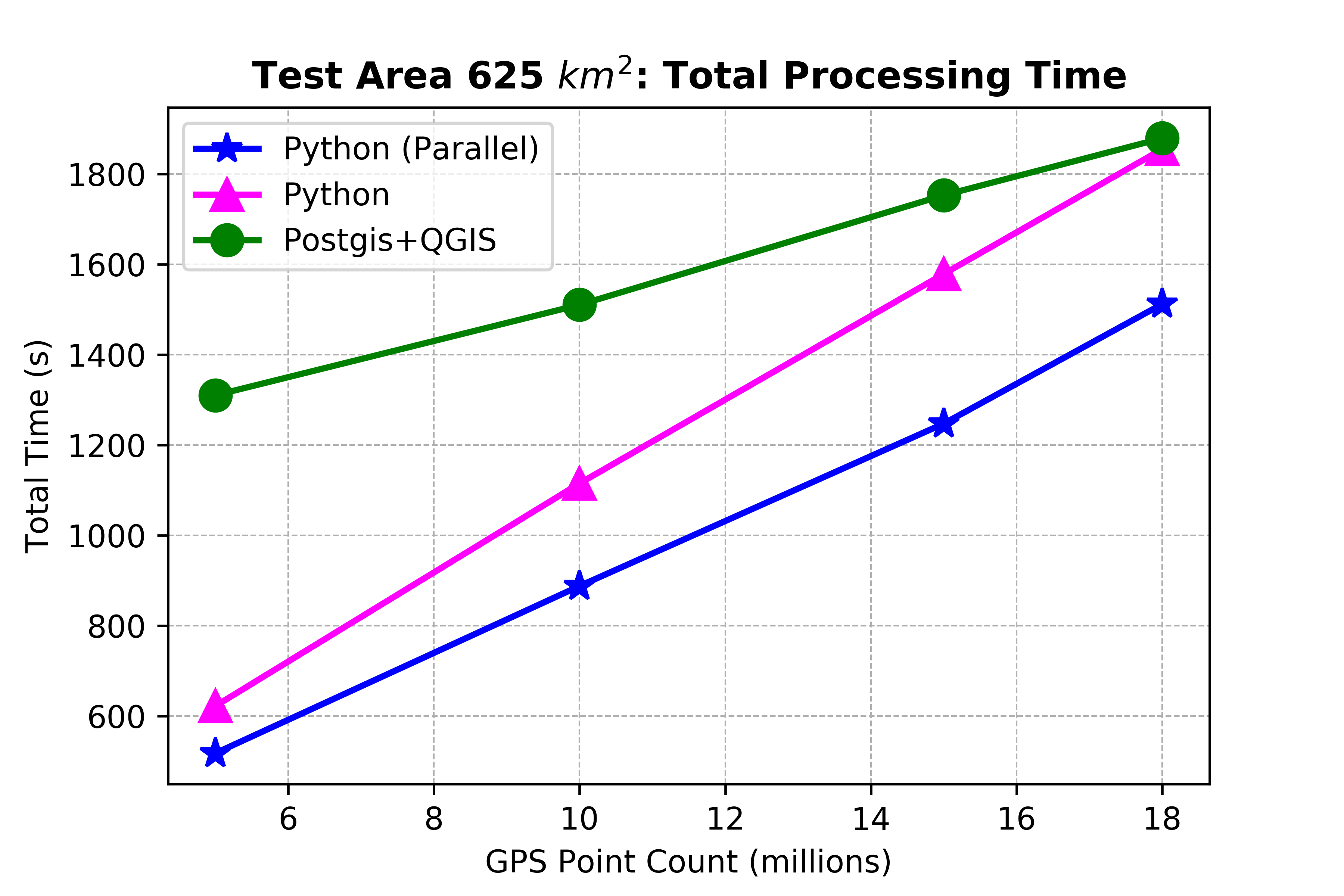}
  \caption{}
  \label{fig:625_all}
\end{subfigure}%
\begin{subfigure}{.5\textwidth}
  \includegraphics[width=.90\linewidth]{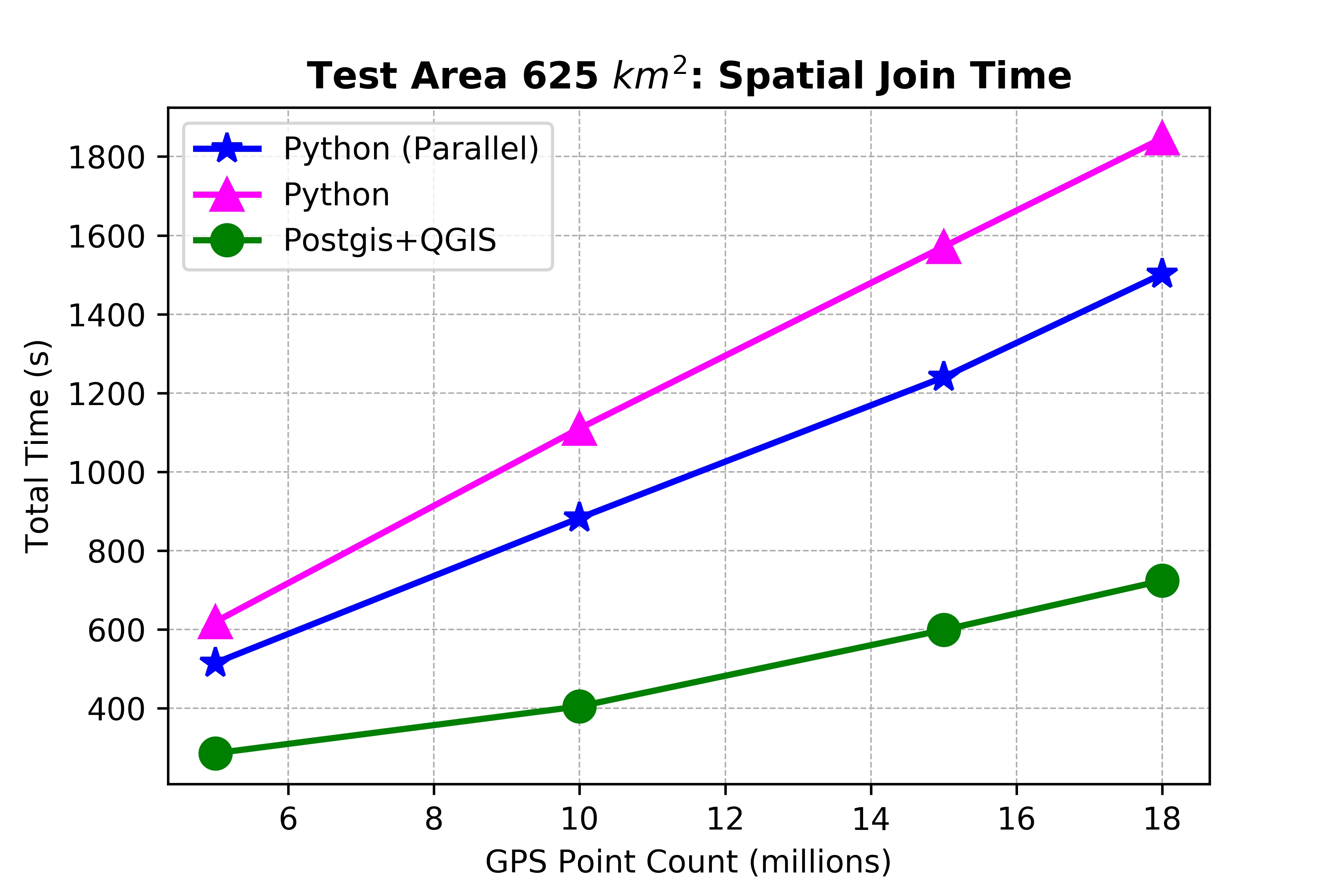}
  \centering
  \caption{}
  \label{fig:s625_all_sptl}
\end{subfigure}
\begin{subfigure}{.5\textwidth}
  \centering
  \includegraphics[width=.90\linewidth]{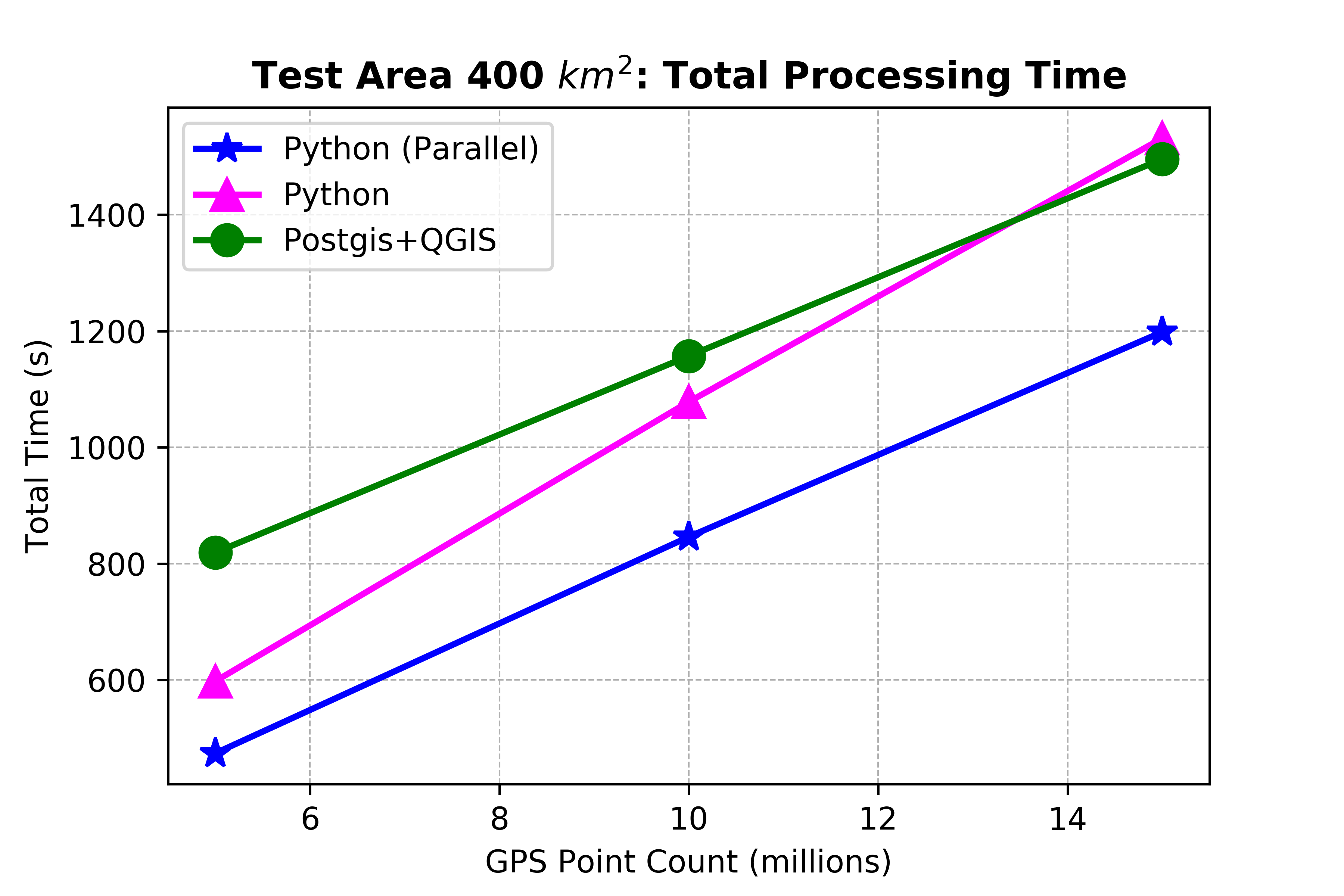}
  \caption{}
  \label{fig:400_all}
\end{subfigure}%
\begin{subfigure}{.5\textwidth}
  \includegraphics[width=.90\linewidth]{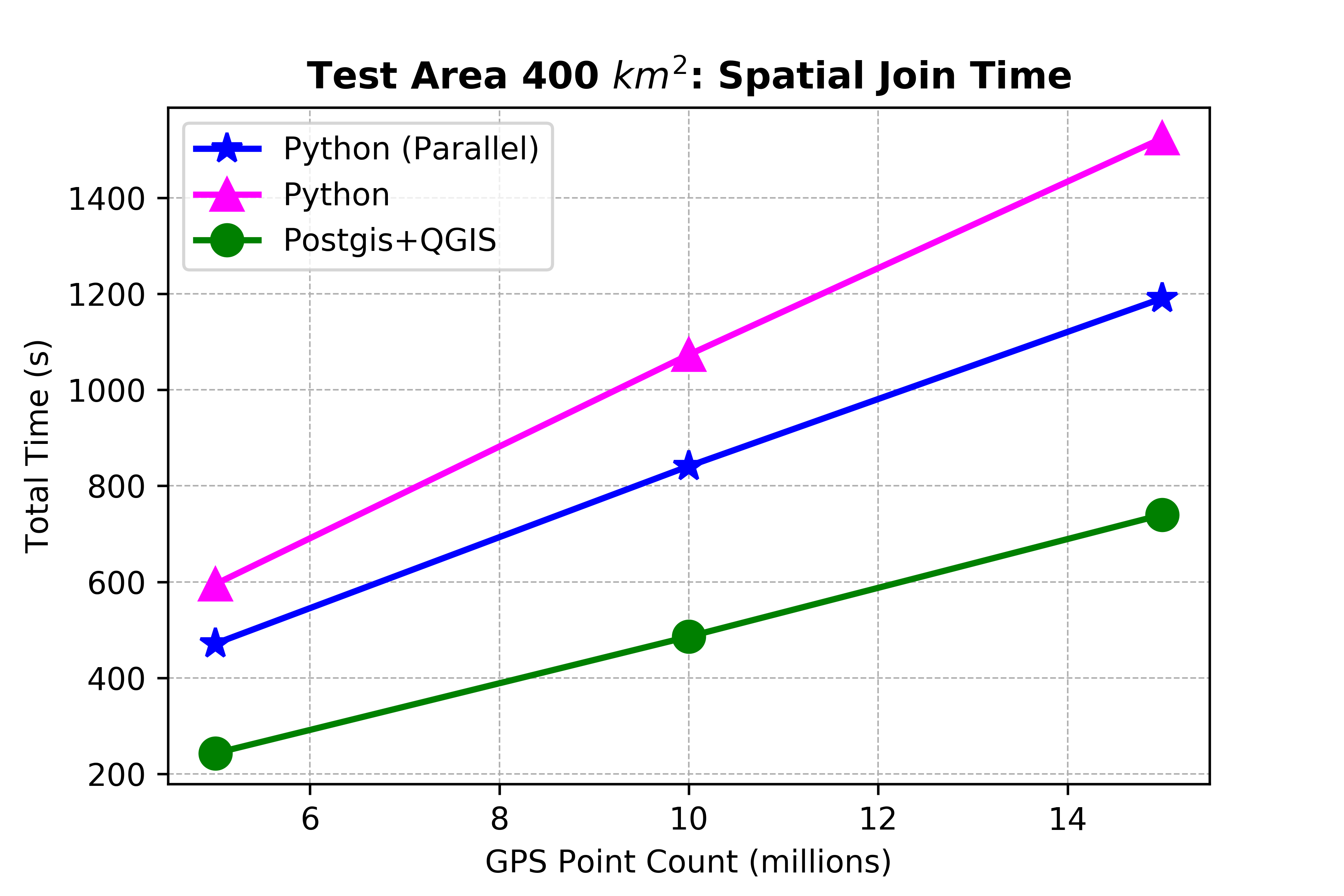}
  \centering
  \caption{}
  \label{fig:s400_all_sptl}
\end{subfigure}
\begin{subfigure}{.5\textwidth}
  \centering
  \includegraphics[width=.90\linewidth]{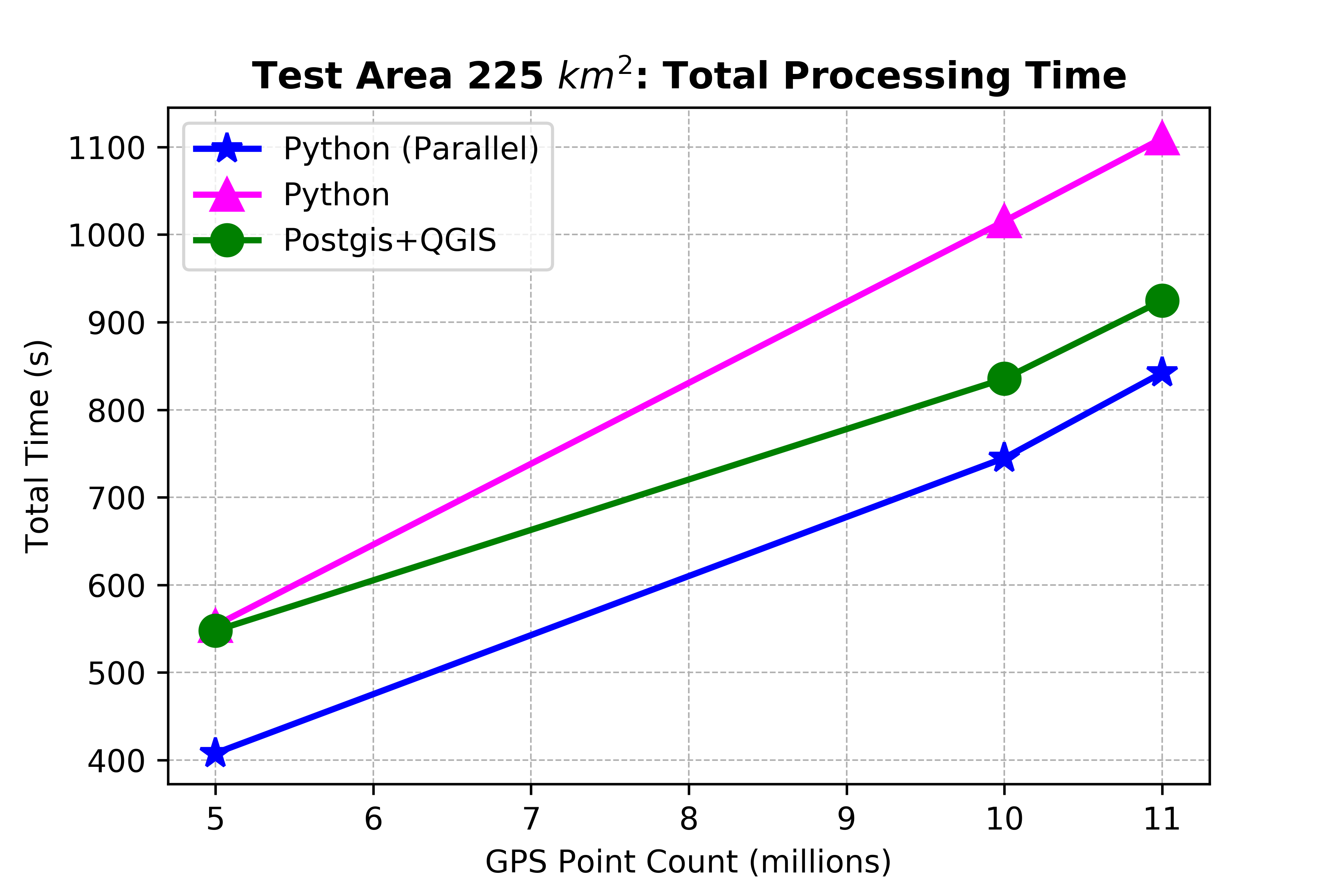}
  \caption{}
  \label{fig:225_all}
\end{subfigure}%
\begin{subfigure}{.5\textwidth}
  \includegraphics[width=.90\linewidth]{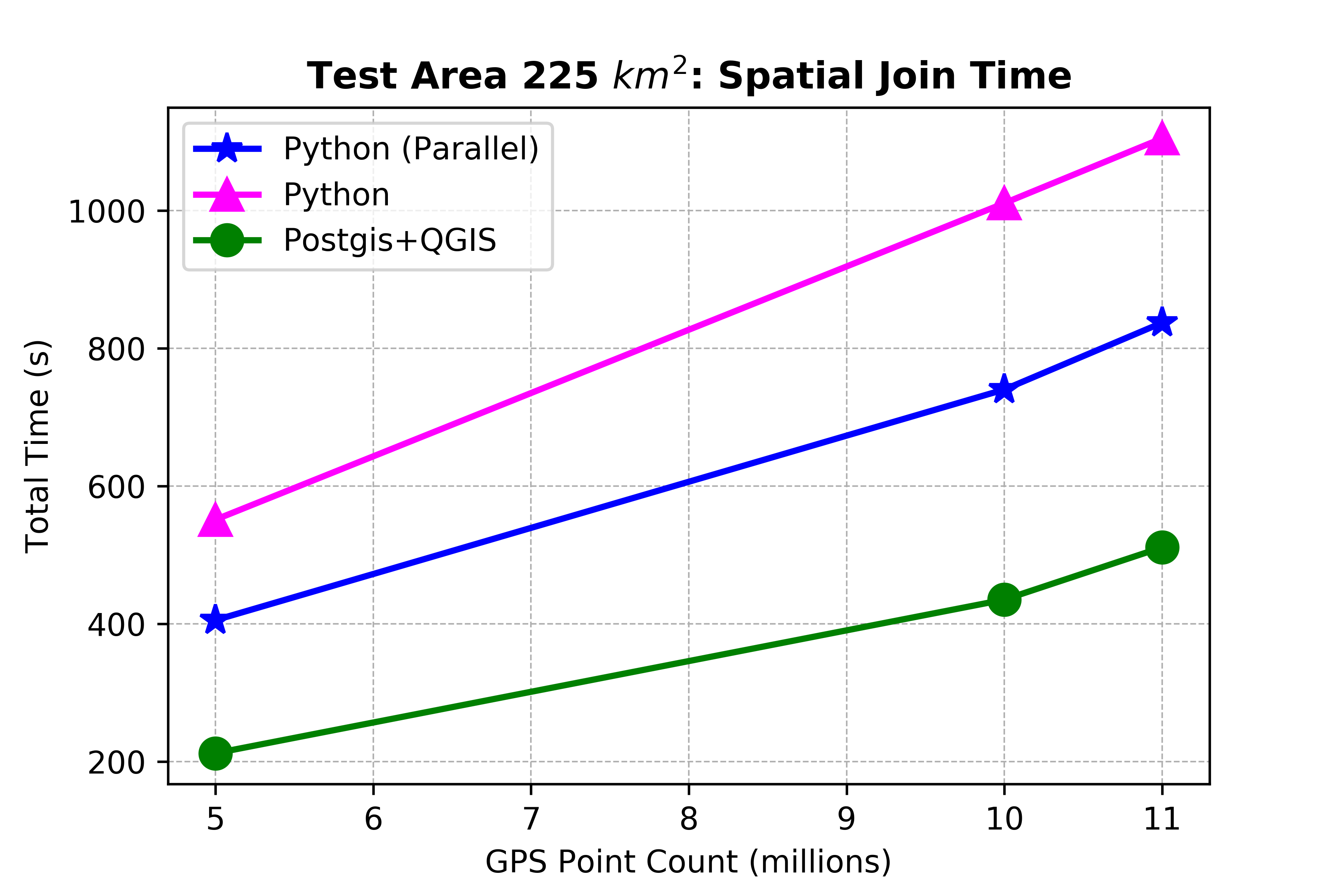}
  \centering
  \caption{}
  \label{fig:s225_all_sptl}
\end{subfigure}
\begin{subfigure}{.5\textwidth}
  \centering
  \includegraphics[width=.90\linewidth]{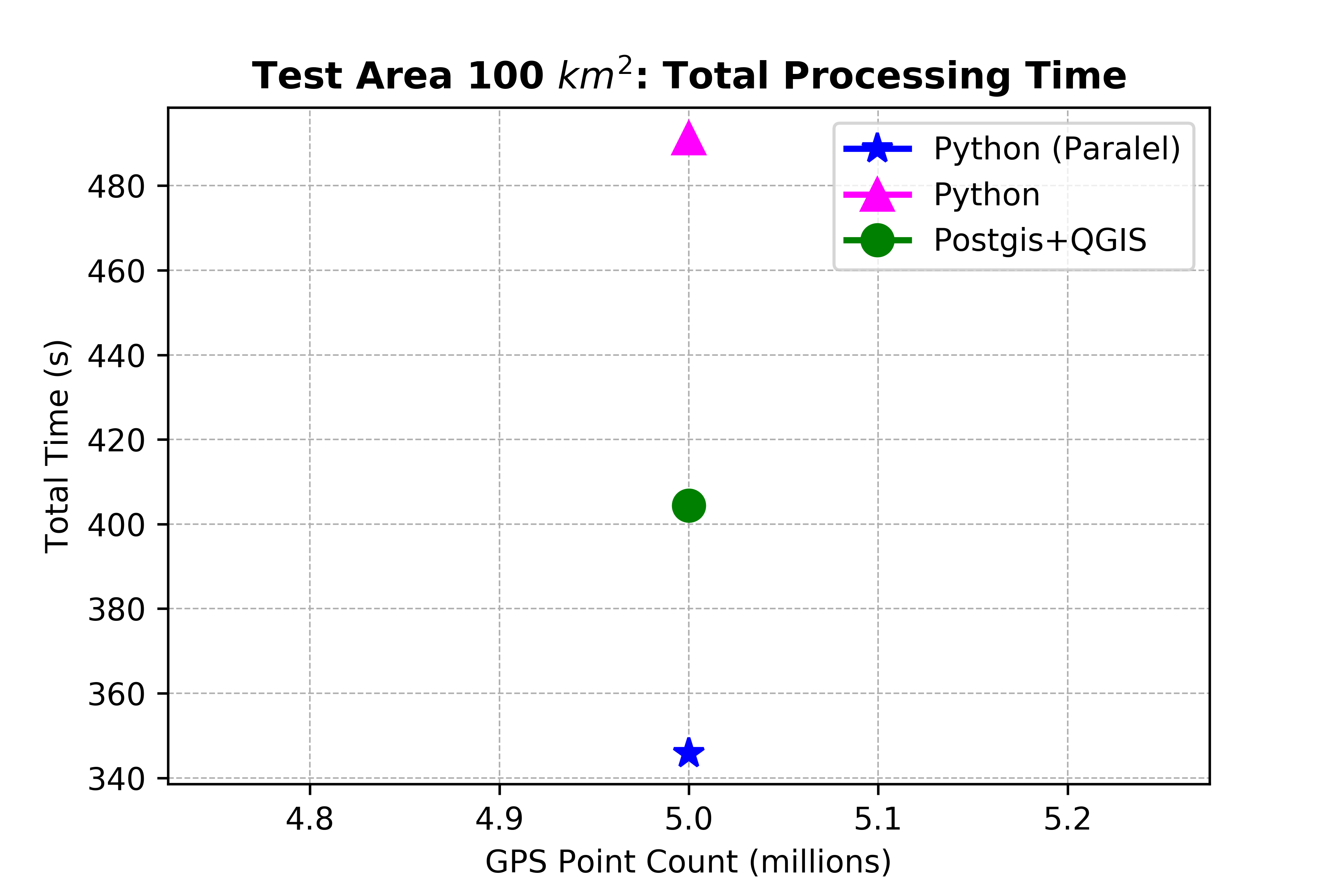}
  \caption{}
  \label{fig:100_all}
\end{subfigure}%
\begin{subfigure}{.5\textwidth}
  \includegraphics[width=.90\linewidth]{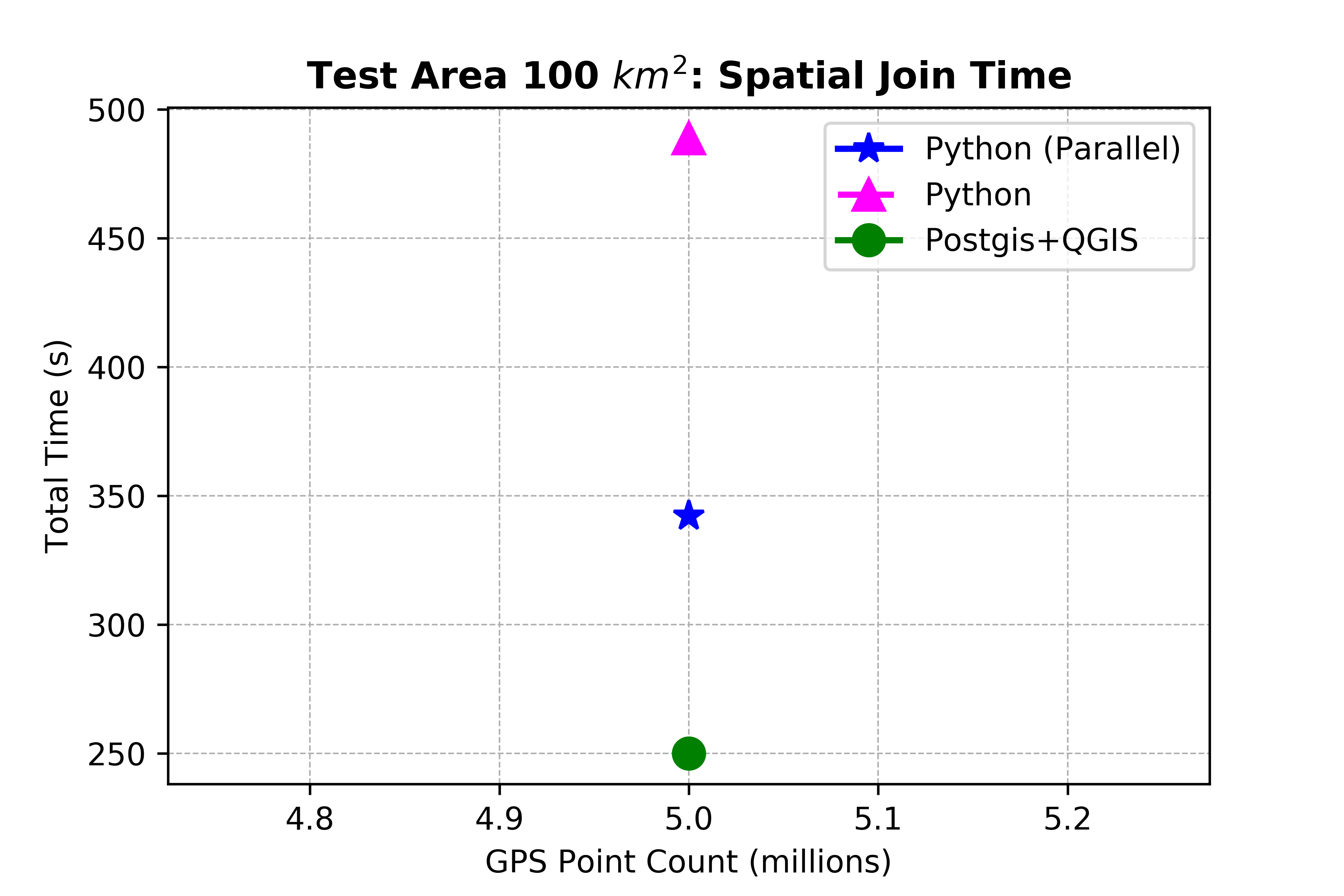}
  \centering
  \caption{}
  \label{fig:s100_all_sptl}
\end{subfigure}
\caption{Comparison of total processing time and spatial join for each test area (QGIS method's results are excluded).}
\label{fig:gps_comparison}
\end{figure}

With the use of the output data from the experiments, comparison plots are created. Figures \ref{fig:gps_comparison_with_qgis} and Figures \ref{fig:gps_comparison} show the performance comparison of methods for each test area. Firstly, as seen in Figure \ref{fig:900_all_qgis}, Python (Parallel) and Python methods achieve best performance in terms of total processing time and followed by PostGIS+QGIS method within the 900 $km^2$ test area. On the other hand, PostGIS+QGIS achieves the best performance for spatial join time measure (Figure \ref{fig:900_all_sptl_qgis}). QGIS method achieves very poor performance for each measure. Because of this, QGIS method is excluded from plots in the Figures \ref{fig:gps_comparison}.

The results for both measures are very similar for the rest of the test areas (Figure \ref{fig:gps_comparison}). It is also visible that the Python method without parallelization is slower than PostGIS+QGIS method in small areas though performance increases with respect to the PostGIS+QGIS method while the test area size increases (Figure \ref{fig:625_all}, \ref{fig:400_all}, \ref{fig:225_all}). In the spatial join measure, PostGIS+QGIS achieved better performance followed by the Python (Parallel) method. With the increase of GPS point count, the difference between Python methods and PostGIS+QGIS method also increases.

%

\begin{figure}[ht]
    \centering
    \includegraphics[width=.6\linewidth]{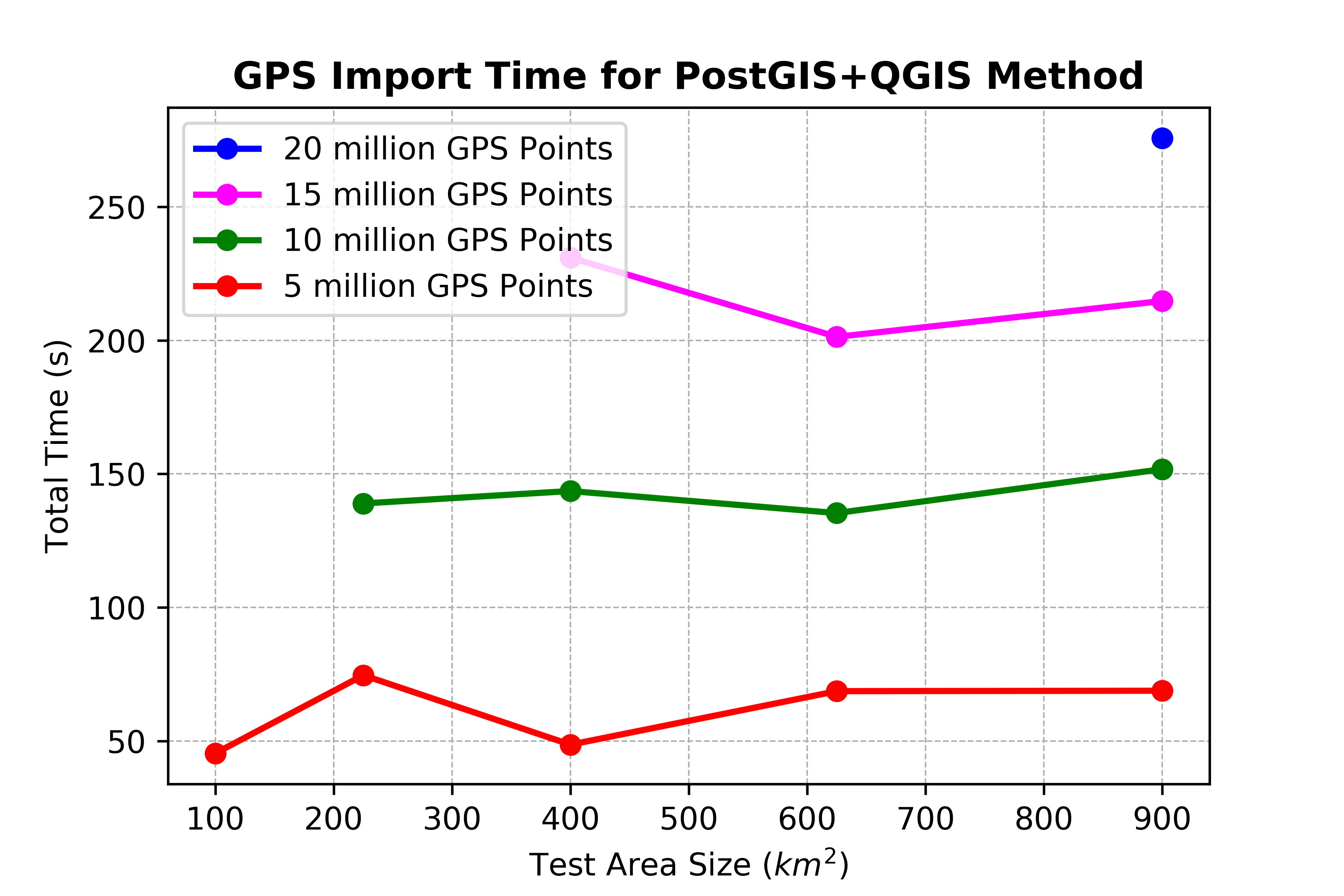}
    \caption{Time spent for GPS trajectory data to database import for PostGIS+QGIS method.}
    \label{fig:gps_import}
\end{figure}

There are two major reasons for the performance decrease of PostGIS+QGIS method in total processing time. The first and most important reason is the requirement of a vector grid. As summarized in Table \ref{tab:grid_creation}, grid creation time is proportional to the test area size and increases as the test area size increases. The second reason is the importing time of the GPS trajectories to the database. Unlike the QGIS method and Python methods, GPS trajectories required to be imported to the database before starting the rest of the process for PostGIS+QGIS method. Figure \ref{fig:gps_import} shows that this process is dependent on the GPS point count and not dependent on test area size.

\begin{table}[H]
\caption{Grid creation time for each test area.}
\label{tab:grid_creation}
\centering
  \begin{tabular}{| c | c |}
    \hline
    \textbf{Test Area Size ($km^2$)} & \textbf{Time (s)}\\
    \hline
    100 & 78 \\
    \hline
    225 & 164 \\
    \hline
    400 & 299 \\
    \hline
    625 & 466 \\
    \hline
    900 & 706 \\
    \hline
  \end{tabular}
\end{table}

{}

\begin{figure}[H]
\captionsetup[subfigure]{justification=centering}
\begin{subfigure}{.5\textwidth}
  \centering
  \includegraphics[width=\linewidth]{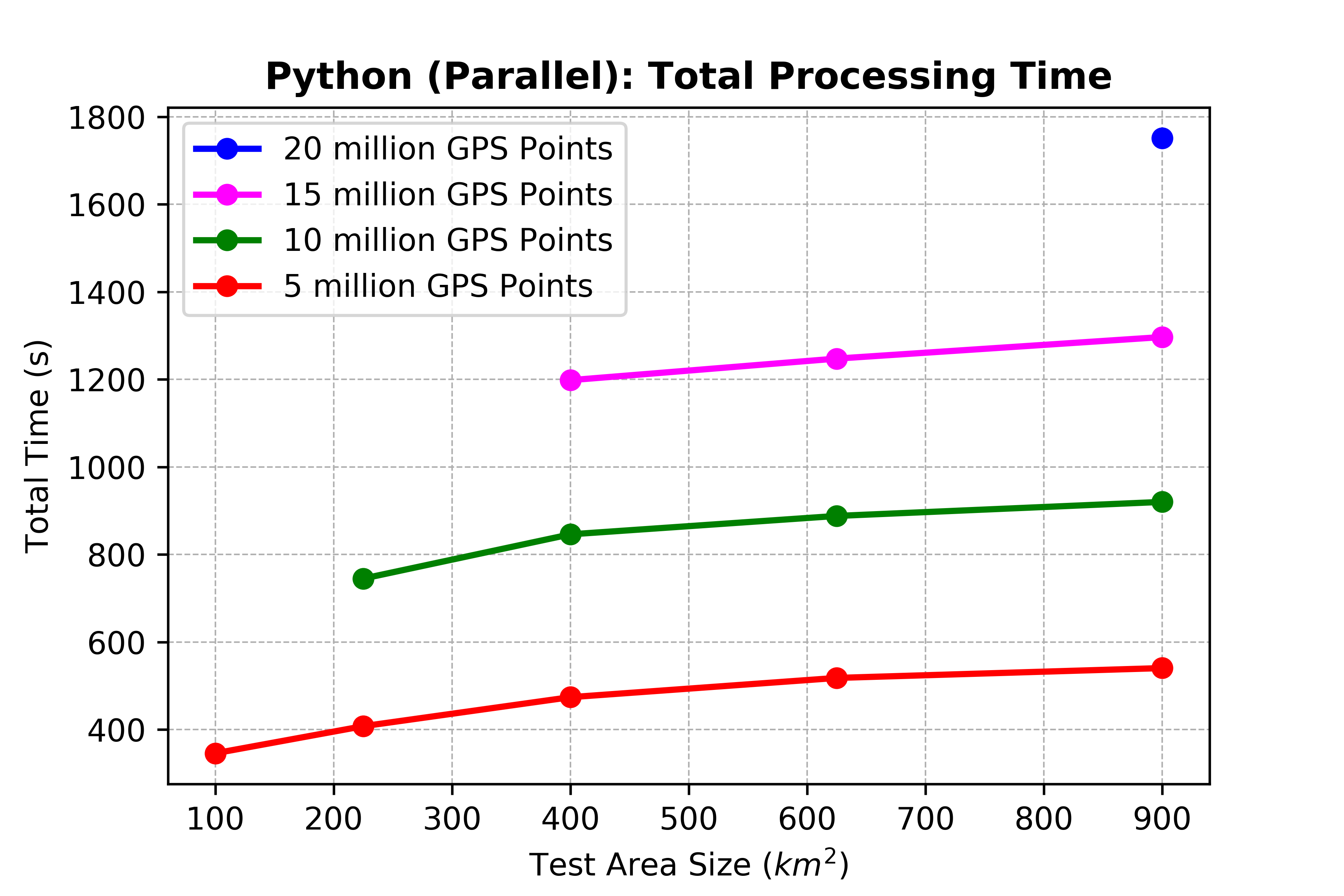}
  \caption{}
  \label{fig:gps_test_area_py_parallel}
\end{subfigure}%
\begin{subfigure}{.5\textwidth}
  \includegraphics[width=\linewidth]{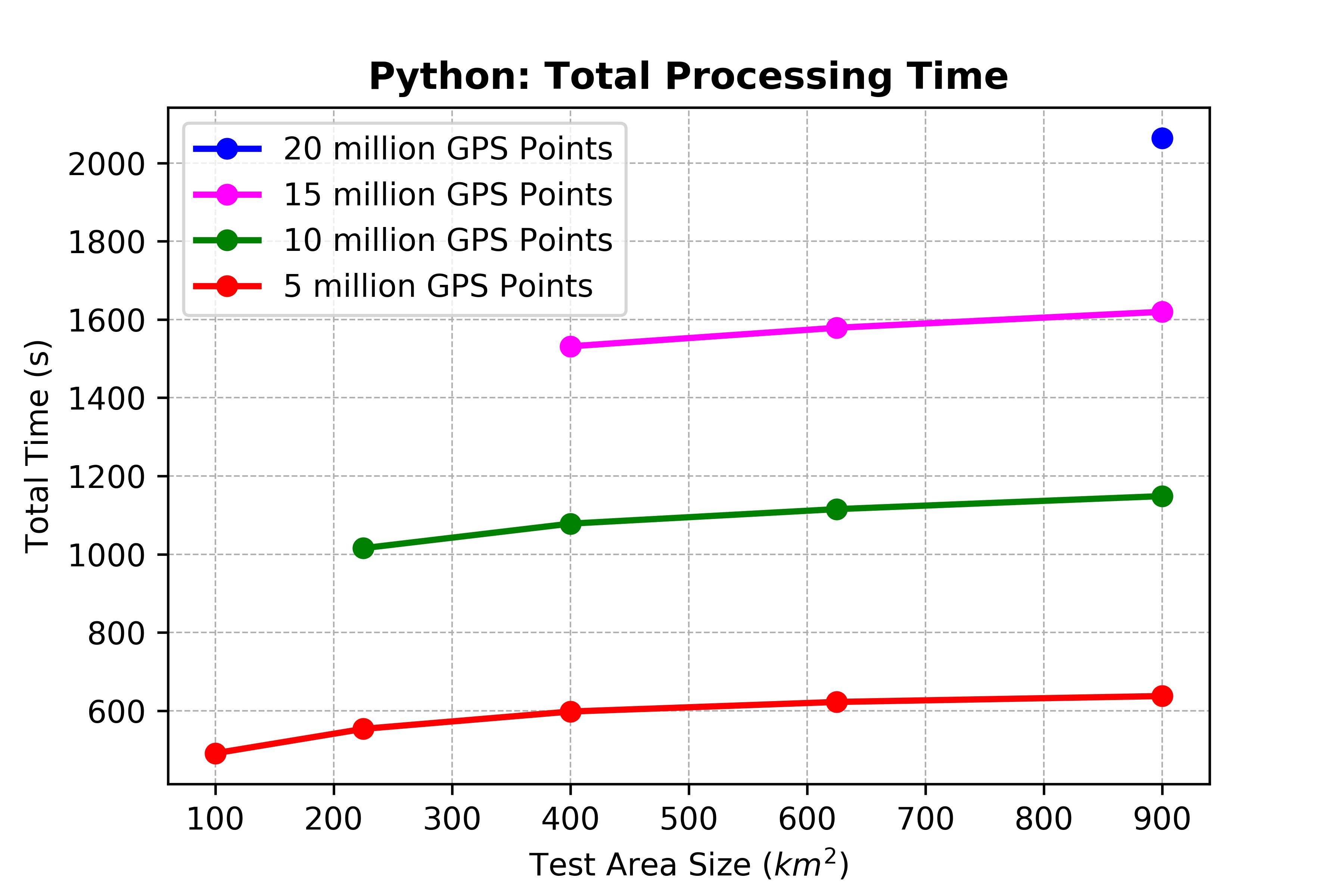}
  \centering
  \caption{}
  \label{fig:gps_test_area_py}
\end{subfigure}
\begin{subfigure}{.5\textwidth}
  \includegraphics[width=\linewidth]{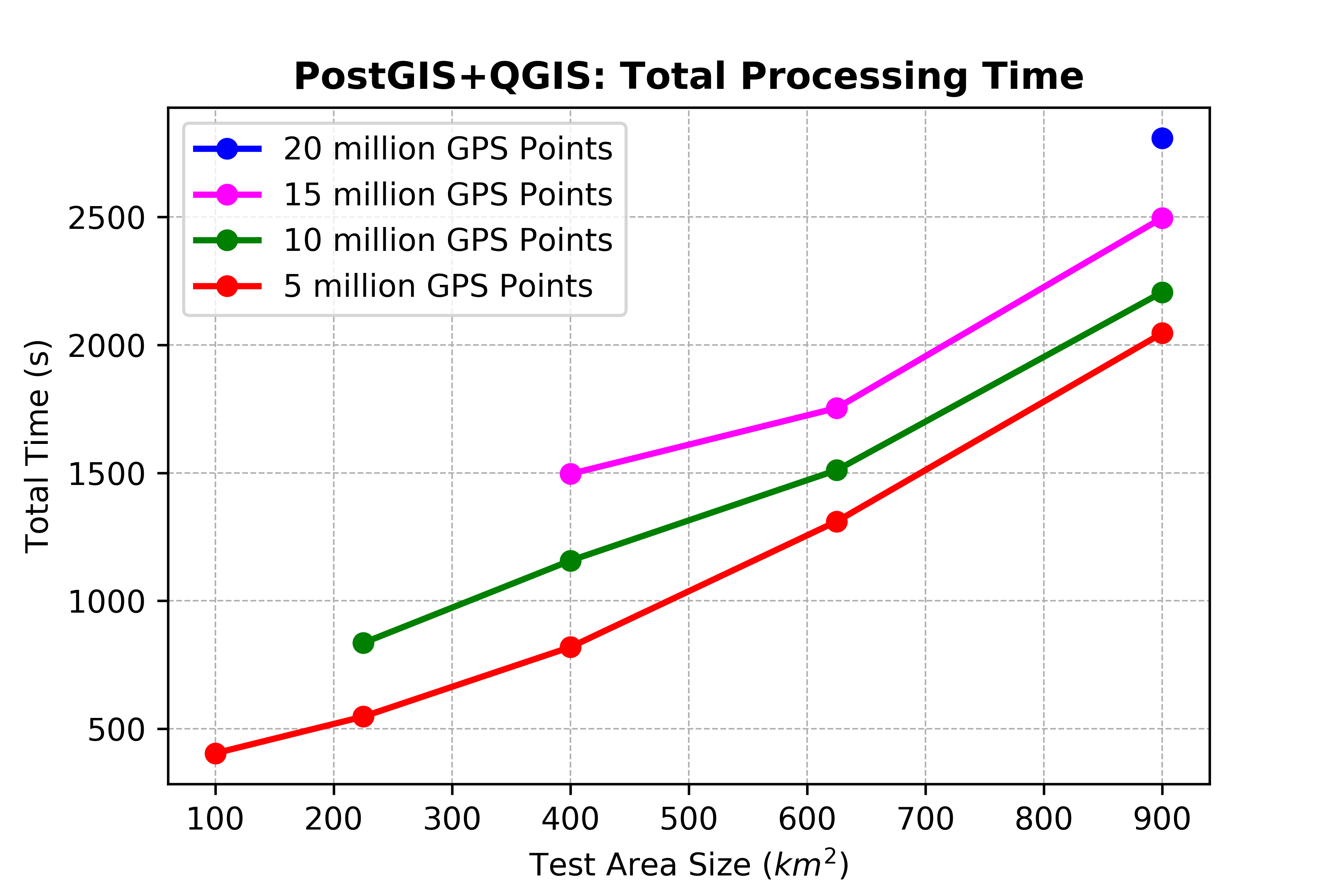}
  \centering
  \caption{}
  \label{fig:gps_test_area_pg_qgis}
\end{subfigure}
\begin{subfigure}{.5\textwidth}
  \includegraphics[width=\linewidth]{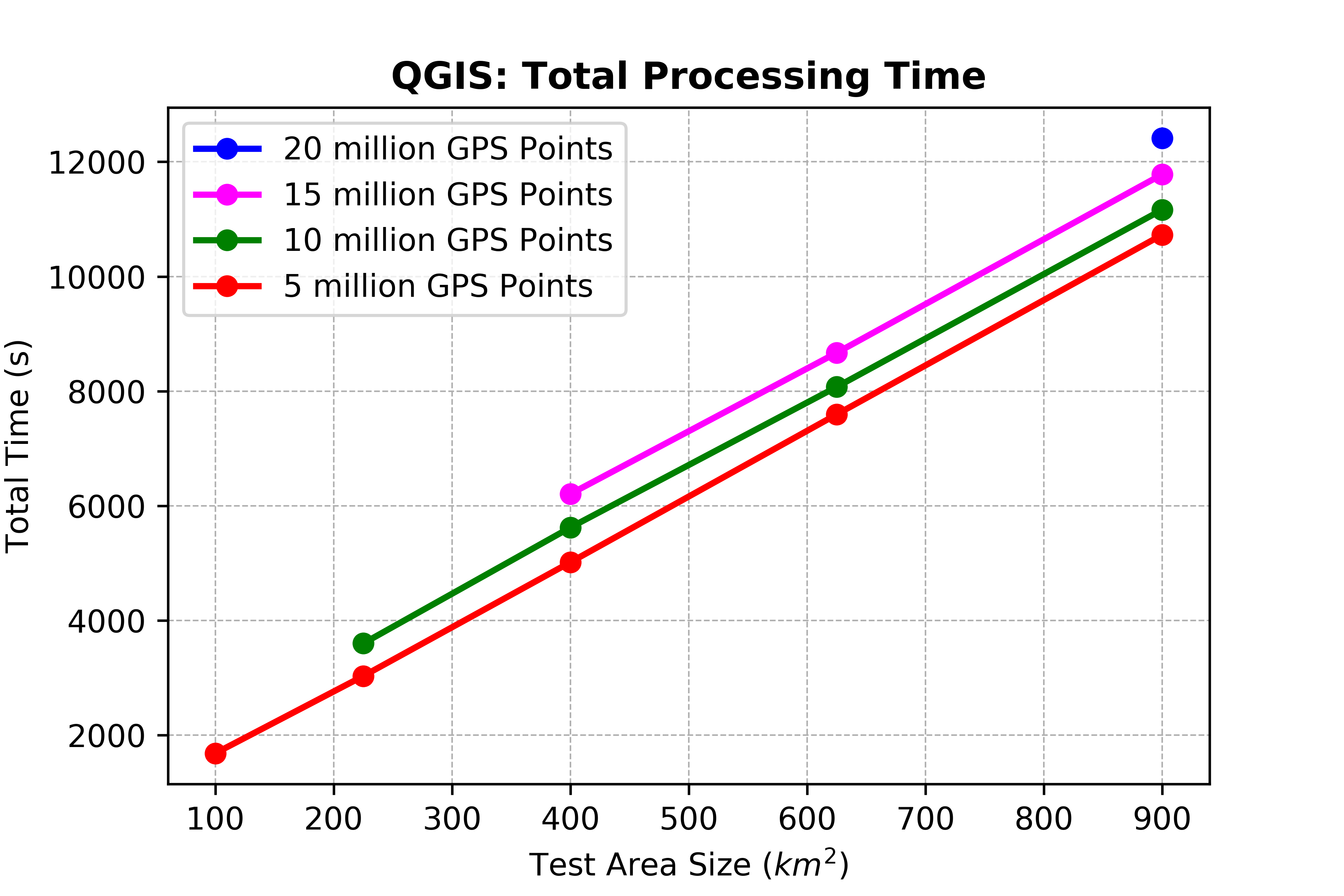}
  \centering
  \caption{}
  \label{fig:gps_test_area_qgis}
\end{subfigure}
\caption{GPS point count vs. test area size comparison of each method; \textbf{(a)} Python (Parallel) \textbf{(b)} Python, \textbf{(c)} PostGIS+QGIS, \textbf{(d)} QGIS.}
\label{fig:methods_comparison}
\end{figure}

Lastly, method results are compared internally with the performance measure by test area size when GPS count is kept constant in plots (Figure \ref{fig:methods_comparison}). As per this comparison, total processing time of the QGIS method increases proportional to the test area size (Figure \ref{fig:gps_test_area_qgis}). Similar to QGIS method, PostGIS+QGIS method's total processing time also increases proportional to test area size though the increase is steeper compared to the QGIS method. On the other hand, both Python (Parallel) and Python methods show very few increases when test area size increases. Their performances are proportional to GPS point count. Python (Parallel) method is faster than Python method.

In addition to the experiments defined in Section \ref{sec:experimental_setup}, an additional experiment was also carried out to understand the limitations of these methods. This experiment was carried out with test area size 22120 $km^2$ and approximately 25 million GPS points around Montreal. Python and Python (Parallel) methods have been able to process and rasterize this area in 40 and 52 minutes respectively. On the other hand, QGIS and PostGIS+QGIS methods couldn't process this larger test area due to grid creation with the current hardware. Since grid creation time is proportional to test area size for both methods, in the case of a big test area, grid creation cannot be possible with the use of QGIS and at the end QGIS crashes.

\section{Conclusions}
\label{sec:conclusions}
This study evaluates the methods for rasterization of GPS trajectories. Evaluation is carried out for QGIS, PostGIS+QGIS methods and our Python and Python (Parallel) implementations. For evaluation, an experiment was carried out with varying test area and GPS trajectory size. Total processing time and spatial join time were adopted as the evaluation metric.

According to the results, the Python (Parallel) method achieves the best results among the compared methods. The Python method also showed better results with respect to QGIS and PostGIS+QGIS methods. PostGIS+QGIS method achieves the best result for spatial join. QGIS shows the worst performance for both of the metrics.

Python and Python (Parallel) methods perform slower than PostGIS+QGIS method for spatial join metric. This issue is a result of the time for indexing operation that our implementation spent which is more than the spatial join operations carried out by PostGIS. Indexing operation only dependent to GPS point data size but PostGIS+ QGIS method performance is dependent both to GPS data size and the test are size. Also, when compared to the spent time for grid creation, this delay caused by indexing is negligible. Moreover, indexing operation is more robust than grid creation. On the contrary grid creation consumes too much memory and prone to crashes. Although the PostGIS+QGIS method achieves the best spatial join performance, due to the disadvantage of grid creation and import time required for GPS points, the total performance decreased very fast while the test area size increased. Grid creation can be considered as one-time cost though it is still a disadvantage for the possible cases of different work areas in different applications domains. Similar to the PostGIS+QGIS method, in addition to weak performance of the spatial join, QGIS also performed worse while the test area size increases.

On the other hand, the Python methods' performance is proportional to point count of the GPS trajectories. This feature is proven with additional experiment which has wider test area. As per results, Python methods can work in large areas though QGIS and PostGIS+QGIS methods fail to achieve this. Because the performance is not dependent to test area size and being suitable to parallelization, it is possible to increase performance of Python methods' with distributing computation into more processor cores and/or computation clusters.

As a conclusion, our implementation performs better than QGIS and PostGIS+QGIS methods and can be used for GPS trajectory rasterization. The use of our implementation is not limited to GPS trajectory rasterization. It is also possible to use our implementation in similar problems which require rasterization and aggregation of big point-based datasets into structured grids such as spatial social media data analysis. In addition, the integration of our implementation would increase its usage in other research domains which benefit from QGIS but requiring better performance. It is possible to integrate our implementation scripts into QGIS since support Python programming language, but further research needed to determine if libraries like Swifter are compatible with QGIS Python environment.

\noindent \textbf{Acknowledgement: }
This preprint has not undergone peer review (when applicable) or any post-submission improvements or corrections. The Version of Record of this article is presented in Computational Science and Its Applications – ICCSA 2021 and published in Lecture Notes in Computer Science, and is available online at \url{https://doi.org/10.1007/978-3-030-86653-2_1}.

\bibliographystyle{splncs04}
\bibliography{references}

\end{document}